\documentclass[aps,prb,reprint,numerical,superscriptaddress]{revtex4-1}

\usepackage[utf8]{inputenc}
\usepackage{amssymb}
\usepackage{amsmath}
\usepackage{multirow}

\usepackage[pdftex]{graphicx}
\usepackage{epstopdf} 
\usepackage{xspace}
\usepackage{dsfont}
\usepackage{bm}
\usepackage[normalem]{ulem}
\usepackage{hyperref}
\hypersetup{
    colorlinks,%
    citecolor=blue,%
    filecolor=blue,%
    linkcolor=blue,%
    urlcolor=blue
}

\providecommand{\ket}[1]{\vert #1\rangle} 
\providecommand{\bra}[1]{\langle #1\vert} 
\providecommand{\mean}[1]{\langle #1 \rangle} 

\usepackage{color}

\begin{document}

\title{Macroscopic transverse drift of long current-induced spin coherence in two-dimensional electron gases}
\author{F. G. G. Hernandez}
 \email[Corresponding author.\\ Electronic address: ]{felixggh@if.usp.br}
\author{S. Ullah}
\affiliation {Instituto de F\'{i}sica, Universidade de S\~{a}o Paulo, S\~{a}o Paulo 05508-090, SP, Brazil}
\author{G. J. Ferreira}
\affiliation {Instituto de F\'{i}sica, Universidade Federal de Uberl\^{a}ndia, Uberl\^{a}ndia 38400-902, MG, Brazil}
\author{N. M. Kawahala}
\author{G. M. Gusev}
\affiliation {Instituto de F\'{i}sica, Universidade de S\~{a}o Paulo, S\~{a}o Paulo 05508-090, SP, Brazil}
\author{A. K. Bakarov}
\affiliation{Institute of Semiconductor Physics and Novosibirsk State University, Novosibirsk 630090, Russia}
\date{\today}

\date{\today}

\begin{abstract}
We imaged the transport of current-induced spin coherence in a two-dimensional electron gas confined in a triple quantum well. Nonlocal Kerr rotation measurements, based on the optical resonant amplification of the electrically-induced polarization, revealed a large spatial variation of the electron g factor and the efficient generation of a current-controlled spin-orbit field in a macroscopic Hall bar device. We observed coherence times in the nanoseconds range transported beyond half-millimeter distances in a direction transverse to the applied electric field. The measured long spin transport length can be explained by two material properties: large mean free path for charge diffusion in clean systems and enhanced spin-orbit coefficients in the triple well.
\end{abstract}

\maketitle

\section{Introduction}

Future electronic technologies based on the spin degree of freedom will require to maintain long quantum coherence times during charge carrier transport in macroscopic devices.\cite{wolf,schliemann} The successful performance of this fundamental task needs focused studies on drift and diffusion in test bench systems as, for example, electron spins in GaAs.\cite{dzhioev,crooker05,cadiz,srep16} One study approach, observed early in bulk samples, is the drift of optically polarized spins by an in-plane electric field imaged by the age of bunches in the spin polarization.\cite{kikkawa99} Beyond the simple acceleration of the electron's charge, the electric field changes the momentum-dependent spin-orbit fields ($B_{SO}$) and manipulates the direction of their spins.\cite{manchon} For two-dimensional electron gases (2DEGs) hosted in a semiconductor quantum well, several reports explored the spin-orbit interaction (SOI) tunability to produce a unidirectional $B_{SO}$ for the diffusive generation of a spin helix.\cite{walser,ishihara} Very recently, the drift in those helical spin systems was also demonstrated showing remarkable properties as the enhancement of the spatial coherence and the electrical current control of the precession frequency.\cite{yang,kunihashi,prl2016} 

A second possibility is the transport study of spins polarized by an electrical current.\cite{vasko,aronov,edelstein,ganichevreview} The generation of in-plane current-induced spin polarization (CISP) have been extensively studied in bulk samples \cite{kato,stern} as well as in p- and n-doped quantum wells.\cite{ganichev1,ganichev2,silov,averkiev,sihnat} The CISP has been associated to the spins spatially homogeneous alignment along $B_{SO}$. In a pioneering work using GaAs epilayers, V. Sih and collaborators studied the drift of electron spins that were polarized in the out-of-plane direction by the spin Hall effect (SHE).\cite{dyakonov,sihprl06} Also in nonlocal experiments, they found that spin currents can be driven over tens of microns in transverse regions with minimal electrical fields, where the transverse spin drift velocities were similar to those for longitudinal charge transport. Furthermore, Y. K. Kato and collaborators also studied the CISP transverse transport in bulk InGaAs using an L-shaped channel.\cite{katoL} Lately, CISP has attracted large attention due to the feasibility to be electrically or optically controlled \cite{stepanov,prb2014} by electron and nuclear spin dynamics.\cite{trowbridge} Nevertheless, the relationship between spin-orbit symmetry and electrical spin generation remains controversial and requires further work.\cite{norman}

Here, we studied the CISP transport in a triple quantum well (TQW) containing a 2DEG. The multilayer system with several subbands occupied offers additional control knobs for the SOI as calculated\cite{esmerindo,khaestkii} and experimentally demonstrated in double quantum wells (DQWs).\cite{prb2014,prb2013} For TQWs, it was predicted that the SOI can be smoothly tuned by the electron occupation, controlled by a gate voltage, and with a contribution arising from the linear Dresselhaus term being stronger than in DQWs.\cite{wangfu} Such structures have been extensively studied by magnetotransport \cite{tqw1,tqw2,tqw3} and also suggested for applications in the production of spin blockers and filters.\cite{tqw4,tqw5} We calculated the inter- and intra-subband spin-orbit coefficients for our sample structure and found a large current-controlled spin-orbit field. We mapped the longitudinal and transverse drift of current-induced spin polarization using space-resolved Kerr rotation (KR) in a macroscopic Hall bar. By the periodic optical control of the CISP,\cite{prb2014} the data revealed transverse transport of spin coherence in the nanoseconds range over millimeter distances opening new paths for spintronic devices.  

\section{Materials}

The TQW sample consists of a 26-nm-thick GaAs central well and two 12-nm lateral wells each separated by 1.4-nm-thick Al$_{0.3}$Ga$_{0.7}$As barriers grown in the [001] direction.  The central well has a larger width in order to be populated because the electron density tends to concentrate mostly in the side wells as result of electron repulsion and confinement. Figure ~\ref{fig:tqw}(a) shows the TQW band structure and subband charge density. The structure was symmetrically delta doped with total electron sheet density n$ _{s}$ = 9.6$\times$10$^{11}$ cm$^{-2}$ and low-temperature mobility $\mu$ = 5$\times$10$^{5}$ cm$^{2}$/Vs. The TQW was embedded in a short-period AlAs/GaAs superlattice in order to shield the doping ionized impurities and efficiently enhance the mobility.\cite{bakarov} The electron density of the central well is about 1.4$\times$10$^{11}$ cm$^{-2}$ and both side wells have approximately equal electron density of 4.1$\times$10$^{11}$ cm$^{-2}$ (see SM-A for the magnetoresistance data).\cite{SM}

\section{Theoretical model}

\subsection{Spin-orbit coefficients}

For systems with more than one subband occupied, a new intersubband-induced spin-orbit interaction was first proposed by J. C. Egues and collaborators in Ref.~\onlinecite{esmerindo}. Here, we will use the notation for the spin-orbit couplings introduced by the same group in Ref.~\onlinecite{calsaverini}. To characterize the TQW we consider an effective mass model for the conduction band within the Hartree approximation, which allows us to calculate the spin-orbit couplings (SOC) self-consistently.\cite{fu} We define the crystallographic directions as follows: $\hat{x}=[110]$, $\hat{y}=[1\overline{1}0]$, and $\hat{z}=[001]$. For an heterostructure confined along $z$, the transversal Hamiltonian reads:

\begin{equation}
 H_z = \dfrac{p_z^2}{2m} + V(z) + V_H[\rho(z)],
\end{equation}
where $m$ is the effective mass, $V(z)$ is the structural confining potential of the heterostructure, and $V_H[\rho(z)]$ is the Hartree potential, which is a functional of the density $\rho(z)$. The planar Hamiltonian, $H_{xy}$, simply gives us the parabolic subband dispersion. The subband $\nu$ energy $\varepsilon_\nu$ and eigenstate $\varphi_\nu(z)$ are obtained solving the Schrödinger and Poisson equations self-consistently. 

The intra- and inter-subband spin-orbit coefficients\cite{fu} can be calculated using the eigenstates of $H_z$ as:
\begin{align}
 \eta_{\nu,\nu'}   &= \bra{\nu} \eta_w V' + \eta_H V'_H \ket{\nu'},\\
 \Gamma_{\nu,\nu'} &= \gamma \bra{\nu} k_z^2 \ket{\nu'},
\end{align}
where $\ket{\nu}$ represents $\varphi_\nu(z)$, $\gamma$ is the bulk Dresselhaus coefficient, $V' = \partial_z V(z)$, $V'_e = \partial_z V_H[\rho(z)]$, and $\eta_w$ and $\eta_H$ are bulk coefficients from the $k\cdot p$ model.\cite{esmerindo,fu,WinklerBook} In the matrix diagonal, the usual Rashba SOC for each subband is $\alpha_\nu = \eta_{\nu,\nu}$, and the linear Dresselhaus SOC is $\beta_{1,\nu} = \Gamma_{\nu,\nu}$. The remaining nondiagonal terms of the $\eta$ and $\Gamma$ matrices represent the intersubband SOC. Finally, the cubic Dresselhaus SOC at the Fermi level is $\beta_{3,\nu} \approx \gamma \pi n_\nu/2$, where $n_\nu$ is subband areal density.

\begin{figure}[ht!]
 \centering
 \includegraphics[width=1\columnwidth,keepaspectratio=true]{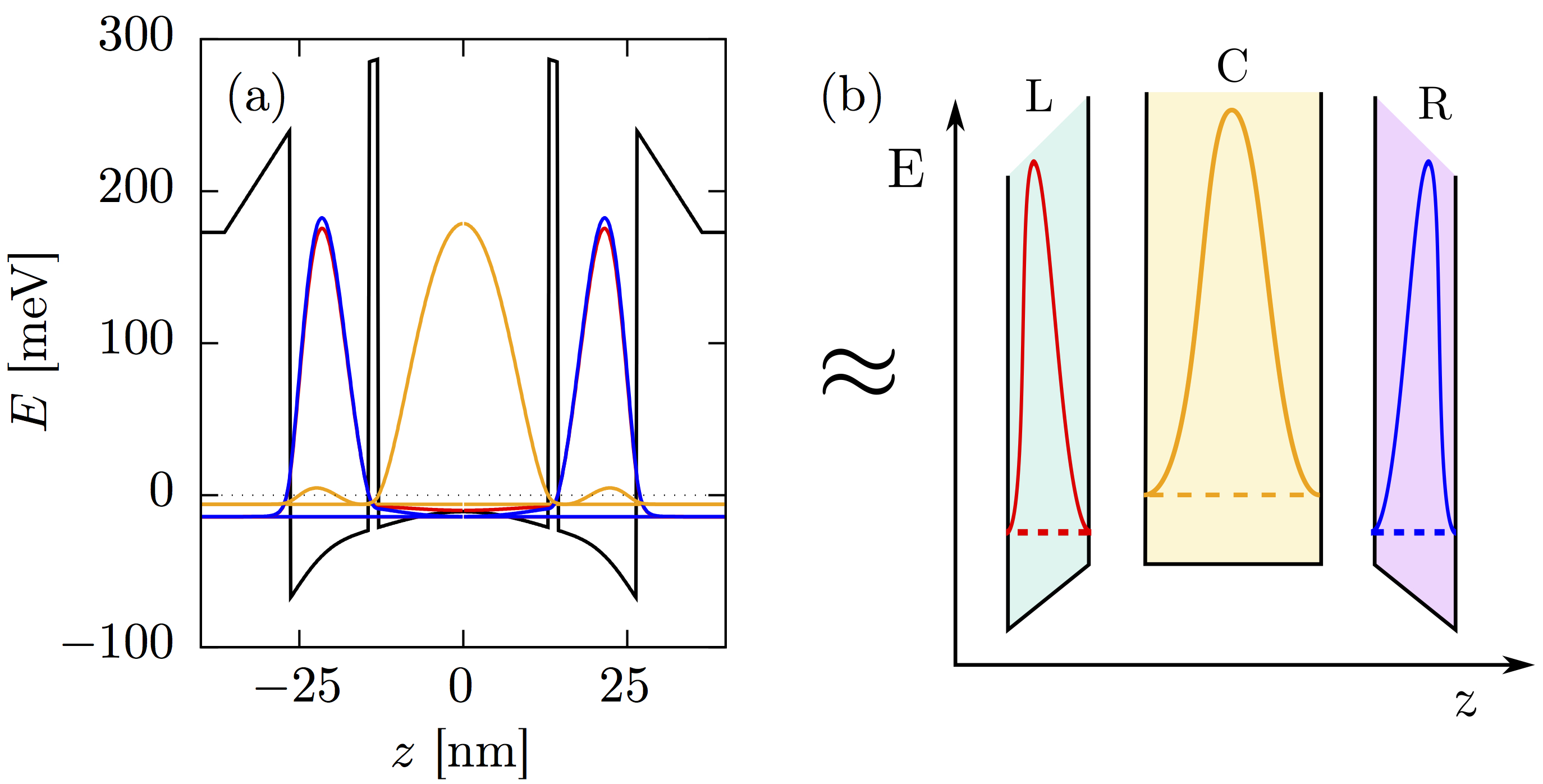}
 \caption{
 (a) Self-consistent solution of Schrödinger and Poisson equations for the triple quantum well. The black lines shows the potential profile and the colored lines show the occupied eigenstates of the first (red), second (blue), and third (orange) subbands.
 (b) Schematic representation of the triple quantum well as a set of left (L), central (C), and right (R) independent quantum wells. Since the first and second subbands are nearly degenerates, we use the approximation $\ket{L} \approx (\ket{1}+\ket{2})/\sqrt{2}$, $\ket{R} = (\ket{1}-\ket{2})/\sqrt{2}$, and $\ket{C} \approx \ket{3}$.
 }
 \label{fig:tqw}
\end{figure}

We choose our sample to set a TQW system where the central well is larger than the side wells as described above. In this condition, the first and second subbands are mostly located at the side wells, while the third subband concentrates in the central well; see Fig.~\ref{fig:tqw}(a). Subbands 1 and 2 are nearly degenerate with an energy difference $\Delta_{21} = \varepsilon_2 - \varepsilon_1 = 0.14$~meV, while the third subband has a larger energy with $\Delta_{32} = 8.11$~meV. For the matrices $\eta$ and $\Gamma$ we find:

\begin{align}
 \eta &= 
 \begin{pmatrix}
  0 & 3.42 & 0\\
  3.42 & 0 & -0.77\\
  0 & -0.77 & 0
 \end{pmatrix}\text{ meV\AA},\\
 \Gamma &= 
 \begin{pmatrix}
  3.49 & 0 & -1.10\\
  0 & 3.69 & 0\\
  -1.10 & 0 & 1.42
 \end{pmatrix}\text{ meV\AA},
\end{align}
and $\beta_{3,\nu} = \{ 0.69, 0.68, 0.29 \}$~meV\AA~ for $\nu = \{1,2,3\}$, respectively. Here we have used the GaAs bulk parameters $\gamma = 11$~eV\AA$^3$, $\eta_w = 3.47$~\AA$^2$ and $\eta_H = 5.28$~\AA$^2$.

The TQW setup is interesting as it enhances the effects of $\beta_{1,\nu}$, diminishes those of $\beta_{3,\nu}$, and decouples the third subband from the first two. First, since the third subband is spatially separated from the other two, the nondiagonal terms in $\eta$ and $\Gamma$ are reduced due to the small overlap between the eigenstates. Combining this with the large $\Delta_{32}$ effectively decouples subband $\nu=3$ from $\nu=\{1, 2\}$. Hence, we can neglect the intersubband SOC terms that connect these subspaces. Second, $\beta_{1,\nu}$ essentially measures the curvature of the eigenstate $\ket{\nu}$. For a single quantum well $\beta_{1,\nu} \propto 1/W^2$, where $W$ is the width of the quantum well. The inner barriers that define the TQW enforces the confinement of each subband into the individual wells, thus reducing the effective width $W$ for each eigenstate, which enhances $\beta_{1,\nu}$. Third, notice that since $\beta_{3,\nu} \propto n_\nu$, for a fixed total density $n_T = \sum_\nu n_\nu$, the sum $\sum_\nu \beta_{3,\nu}$ is also constant. Therefore spreading the total density $n_T$ into multiple subbands reduces the value of each $\beta_{3,\nu}$.

\subsection{Effective model for the 2DEG}

To obtain an effective model for the 2DEG,\cite{fu} one projects the full Hamiltonian $H_z + H_{xy} + H_{SO}$ into the subspace spanned by the relevant subbands $\ket{\nu}$. Here we consider $\nu = \{1, 2, 3\}$, noticing that subbands $\nu = 1$ and $\nu = 2$ are nearly degenerate and correspond to a pair of symmetric and anti-symmetric eigenstates, respectively. Therefore we can take the approximation $\Delta_{21} \rightarrow 0$ and rotate the $\nu=\{1, 2\}$ subspace into left (L) and right (R) eigenstates, i.e., $\ket{L(R)} = (\ket{1}\pm\ket{2})/\sqrt{2}$. For convenience, let's refer to the third subband as the central (C) eigenstate, $\ket{C} = \ket{3}$. Within this new LCR basis set ordered as $\nu = \{L, C, R\}$ we have:

\begin{equation}
 H_{xy}^* = 
 \begin{pmatrix}
  H^*_L & 0 & 0\\
  0 & H^*_C & 0\\
  0 & 0 & H^*_R
 \end{pmatrix},
\end{equation}
where $H_\nu^* = \varepsilon_\nu + \frac{1}{2}g\mu_B \bm{B}_\nu(\bm{k})\cdot\bm{\sigma}$, $g$ is the g factor, $\mu_B$ is Bohr magnetron, and $\bm{B}_\nu(\bm{k})$ is the spin-orbit field of subband $\nu$, which reads:

\begin{equation}
 \bm{B}_\nu(\bm{k}) = 
 \begin{pmatrix}
  \left[+\alpha_\nu + \beta_{1,\nu} + 2\beta_{3,\nu}\dfrac{k_x^2-k_y^2}{k_F^2}\right]k_y\\
  \left[-\alpha_\nu + \beta_{1,\nu} - 2\beta_{3,\nu}\dfrac{k_x^2-k_y^2}{k_F^2}\right]k_x
 \end{pmatrix}.
\end{equation}

Within the LCR basis,  $\varepsilon_L = \varepsilon_R = 0$, $\varepsilon_C = \Delta_{32}$, $\alpha_L = -\alpha_R = \eta_{12} \approx 3.5$~meV\AA, $\alpha_C = 0$, $\beta_{1,L} = \beta_{1,R} = \Gamma_{1,1} \approx \Gamma_{2,2} \approx 3.6$~meV\AA, $\beta_{1,C} =1.42$~meV\AA, $\beta_{3,L} = \beta_{3,R} \approx 0.68$~meV\AA, and $\beta_{3,C} = 0.29$~meV\AA. This $H_{xy}^*$ can be understood as the decoupled quantum wells shown in Fig.~\ref{fig:tqw}(b), where the $\ket{L}$ and $\ket{R}$ belong to the effective triangular side wells, while $\ket{C}$ is on the central square well. Interestingly, the L and R states are close to the PSH$^\pm$ regimes, $\alpha_\nu = \pm (\beta_{1,\nu} - \beta_{3,\nu})$, respectively, constituting the crossed PSH regime of Ref.~\onlinecite{fuarxiv}.

\subsection{Drift induced spin-orbit field}

An in-plane electric field $\bm{E}$ shifts the Fermi circle of the electron's parabolic energy dispersion, inducing a drift velocity $\bm{v_d} = \hbar \bm{k}_d/m^* = \mu\bm{E}$. Here $\mu$ is the mobility, $m^*$ is the effective mass, and $\bm{k}_d$ is the shift of the Fermi circle in k-space. Due to random scattering events, the resulting diffusive motion\cite{yangtheo,prl2016,prb2016} allows the electrons to visit all k-points of this shifted Fermi circle. If the scattering time is short compared with spin precession, at every instant the electrons feel an average effective spin-orbit field $\mean{\bm{B}_\nu}$ over the shifted Fermi circle for each subband $\nu$. For an electric field along $y$, $\bm{v}_d = v_d\hat{y}$ and the average field is:

\begin{equation}
 \mean{\bm{B}_\nu} = \dfrac{2m}{\hbar g\mu_B} v_d (\alpha_\nu+\beta_{1,\nu}-2\beta_{3,\nu}) \hat{x}.
\end{equation}

\begin{figure}[ht!]
 \centering
 \includegraphics[width=\columnwidth,keepaspectratio=true]{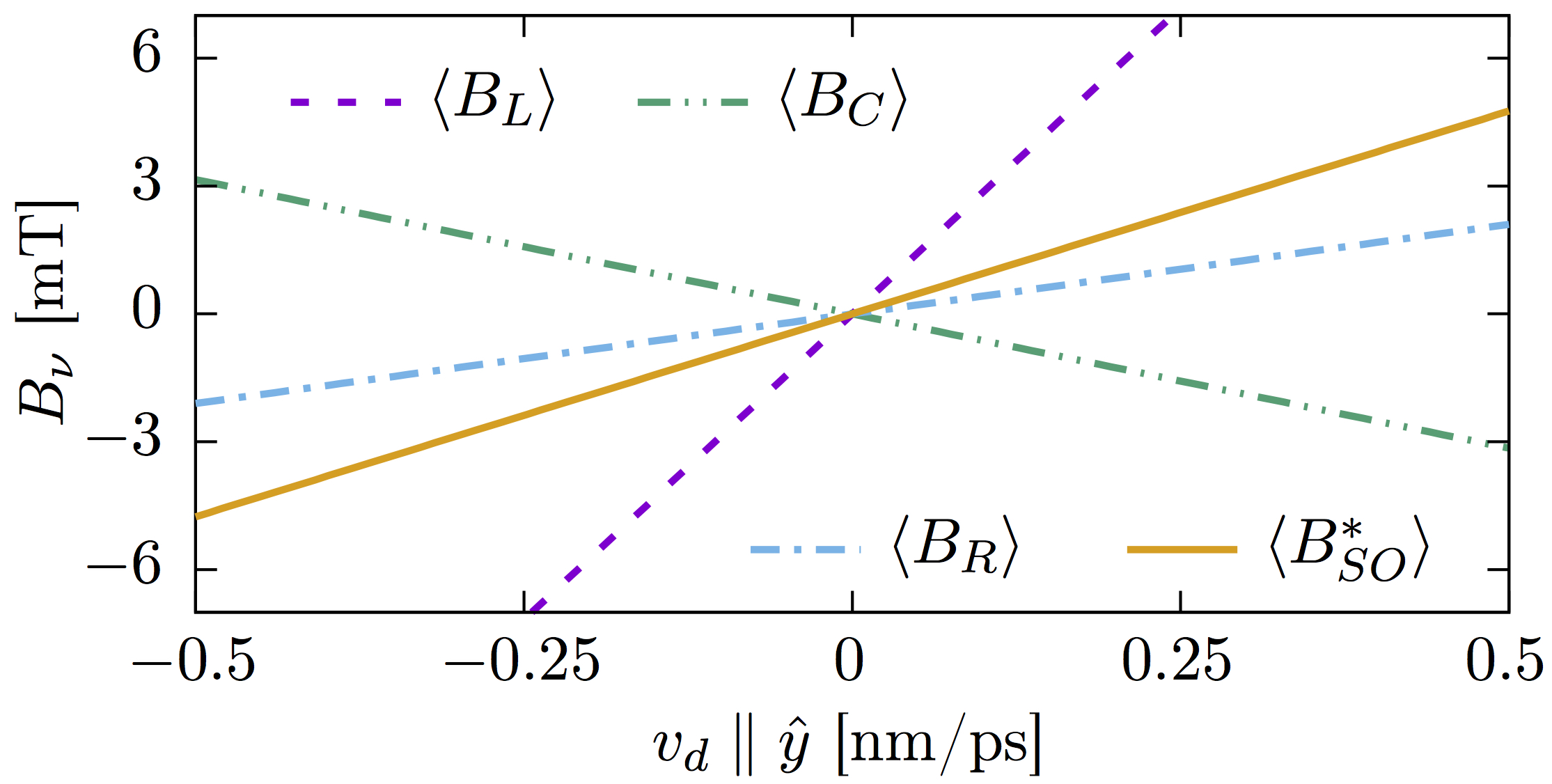}
 \caption{Drift induced spin-orbit fields $\mean{\bm{B}_\nu}$ for each subband $\nu=\{L, C, R\}$ of the effective single wells of Fig.~\ref{fig:tqw}, and the effective $\mean{\bm{B}^*_{SO}}$ averaged over the subband Fermi circles. The drift velocity is along $\hat{y}$ and the spin-orbit fields are along $\hat{x}$.
 }
 \label{fig:Bso}
\end{figure}

Moreover, since the scattering events may induce intersubband transitions, one must also average over the subbands to obtain the final effective spin-orbit field:

\begin{equation}
 \mean{\bm{B}_{SO}^*} = \dfrac{1}{N}\sum_{\nu=1}^{N} \mean{\bm{B}_\nu},
\end{equation}

where the sum runs over the occupied subbands $\nu = 1, 2, ..., N$. The effective spin-orbit fields are shown in Fig.~\ref{fig:Bso} as a function of $v_d$ for the parameters of our sample; see Fig.~\ref{fig:tqw}. Numerically, we find $B_{SO}^* \approx 8.8 v_d$. For a small $v_d = 0.5$~nm/ps we have already $B_{SO} = 4.4$~mT.
 
Similarly if the electric field is set along $\hat{x}$, such that $\bm{v}_d = v_d\hat{x}$, we get $\mean{\bm{B}_\nu} = \frac{2m}{\hbar g\mu_B} v_d(-\alpha_\nu+\beta_{1,\nu}-2\beta_{3,\nu})\hat{y}$. Since the degenerate subbands $\nu = L$ and $R$ have $\alpha_L = -\alpha_R$, the intensities $\mean{B_\nu}$ are equivalent to those of Fig.~\ref{fig:Bso} with $L$ and $R$ switched. These Rashba coefficients cancel out in the sum of the effective field $\mean{\bm{B}^*_{SO}} \parallel \hat{y}$ and its intensity remains the same as shown in Fig.~\ref{fig:Bso}.

\section{Experimental Results}

The TQW sample was patterned in a macroscopic Hall bar with a width $w$ = 200 $\mu$m, length separation (in the $y$ axis) between the side probes $L$ = 500 $\mu$m, and 15 $\mu$m wide bridges connecting the main channel to the side regions as sketched in Fig.~\ref{fig3}(a). 

\subsection{Time-resolved spin dynamics}

\begin{figure}[h!]
\includegraphics[width=1\columnwidth]{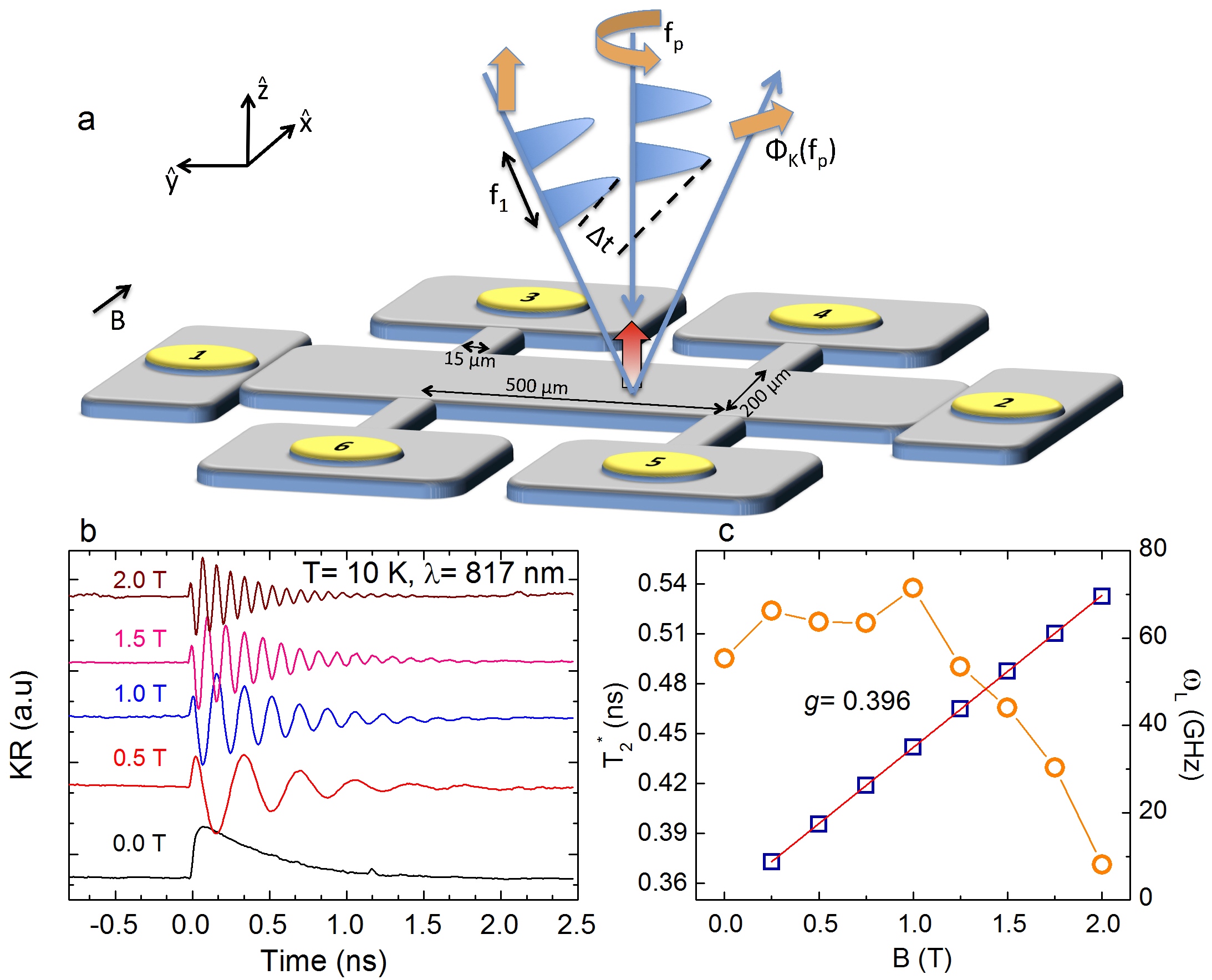}
\caption {Optically-induced spin dynamics. (a) Scheme of the time-resolved KR in the Voigt geometry. (b) KR as function of $\Delta t$ for different B. (c) Magnetic field dependence of the Larmor frequency (squares) and ensemble spin coherence time (circles).}
 \label{fig3}
\end{figure}

\begin{figure}[h!]
\includegraphics[width=1\columnwidth]{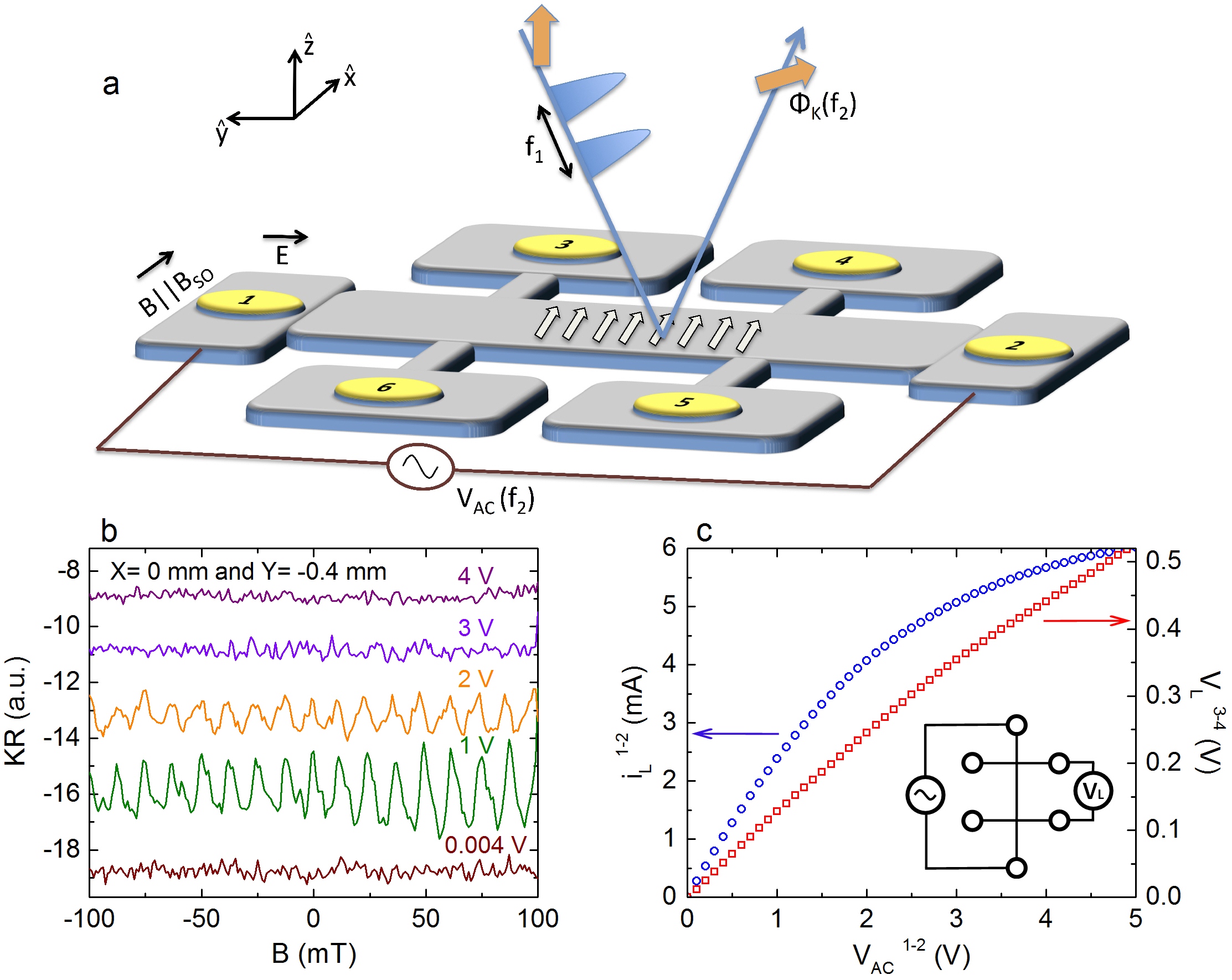}
\caption {Current-induced spin polarization - Longitudinal configuration. (a) Experimental geometry for the optical amplification of the CISP. (b) KR as function of $B$ measured for several $V_{AC}$. (c) Local current and voltage across the sample as a function of the applied voltage. T = 5 K.}
 \label{figquatro}
\end{figure}

First, we studied the sample without the application of electric fields in order to determine the Land\'{e} $g$-factor at $\bm{v_d}=$ 0. We measured the electron spin dynamics using time-resolved KR in the Voigt geometry. We employed a tunable laser with pulse duration of 100 fs and repetition rate of $f_{1}=$ 76 MHz. The spin polarization is generated by a circularly polarized pump and its precession in a transverse magnetic field ($B$) was recorded by a linearly polarized probe laser. The pump beam polarization was modulated at a frequency $f_{p}=$ 50 kHz for detection reference of the Kerr angle ($\phi_K$). Both pulses were focused to approximately 20 $\mu$m. 

Figure~\ref{fig3} shows the KR as function of the time delay ($\Delta t$) between pump and probe pulses for several $B$ at pump/probe power of 1 mW/300 $\mu$W. In Fig.~\ref{fig3}(c), fitting the data with a typical oscillatory exponential decay, we extracted the ensemble spin coherence time ($T_{2}^{*}$) and the Larmor frequency ($\omega_L$). The linear dependence $\omega_{L}=g\mu_{B}B/\hbar$, where $\mu_{B}$ is  Bohr magneton and $\hbar$ is the reduced Planck's constant, gives the Land\'{e} $g$-factor (absolute value) $g$= 0.396 (see SM-B).\cite{SM} 

Furthermore, we found that $T_{2}^{*}=$ 0.5 ns remains constant up to 1 T and then rapidly decreases due to the ensemble spread of the $g$ factor.\cite{zhukov}. This time scale is limited by the spin-orbit coefficients, Rashba and Dresselhaus linear and cubic, through the Dyakonov-Perel spin relaxation mechanism which is dominant in 2DEGs.\cite{walser,kainz,ullah}

\subsection{Longitudinal spin transport}

Now, we switched from optically-induced spin polarization to current-induced spin polarization. Figure~\ref{figquatro}(a) shows the experimental geometry with ${\bm{E}}\perp{\bm{B}}$ and ${\bm{B_{SO}}}\parallel\bm{B}$. We replaced the optical pump pulse by an AC voltage with tunable rms amplitude $V_{AC}$ and fixed modulation frequency $f_{2}=$ 1.1402 kHz for $\phi_K$ lock-in detection. The probe pulse was kept at the same power, wavelength, and focus used in the previous section for maximum signal (see SM-C).\cite{SM} 

In the longitudinal configuration, we tested the CISP response by applying $V_{AC}$ in ohmic contacts at the central channel (1-2) and measuring the KR as a function of $B$. Figure~\ref{figquatro}(b) shows the amplification of the KR at certain resonant fields due to the constructive interference of the CISP dynamics when it is controlled by optical periodic excitation. As sketched, the electron spins in the 2DEG drift parallel to the in-plane $\bm{E}$ and feel a k-dependent spin-orbit field. The spin polarization becomes aligned along $\bm{B_{SO}}$ in a direction perpendicular to $\bm{E}$. The optical pulse train then hits the sample along the out-of-plane direction and rotates the spin polarization towards that direction (detected by polar Kerr rotation). In a time scale faster than the kHz voltage modulation, the precession (with a GHz frequency) of those electrically-polarized spins is amplified by the pulse train if the ensemble conherence time is long enough to persist between pulses with MHz repetition frequency. Such resonant spin amplification (RSA) of the CISP follows the condition $\Delta B=(hf_1)/g\mu_{B}$, and it was previously reported on double quantum well samples.\cite{prb2014} While larger voltages enhance the CISP amplitude, very high voltages are detrimental to the spin coherence due to heating effects. The largest $V_{AC}$ retaining the formation of the RSA pattern thus depends on the electrons temperature (see SM-C for similar data at 1.2 and 10 K).\cite{SM} Figure~\ref{figquatro}(c) shows a mA current flow while the voltage across the sample is only about 10\% of $V_{AC}$ due to the high mobility. The local resistance $R_L=V_L/i_L$ increases linearly with $V_{AC}$ from 50.35 $\Omega$ at 0.5 V to 87.71 $\Omega$ at 5 V.

\subsection{Transverse spin transport}

\begin{figure*}
\includegraphics[width=\textwidth]{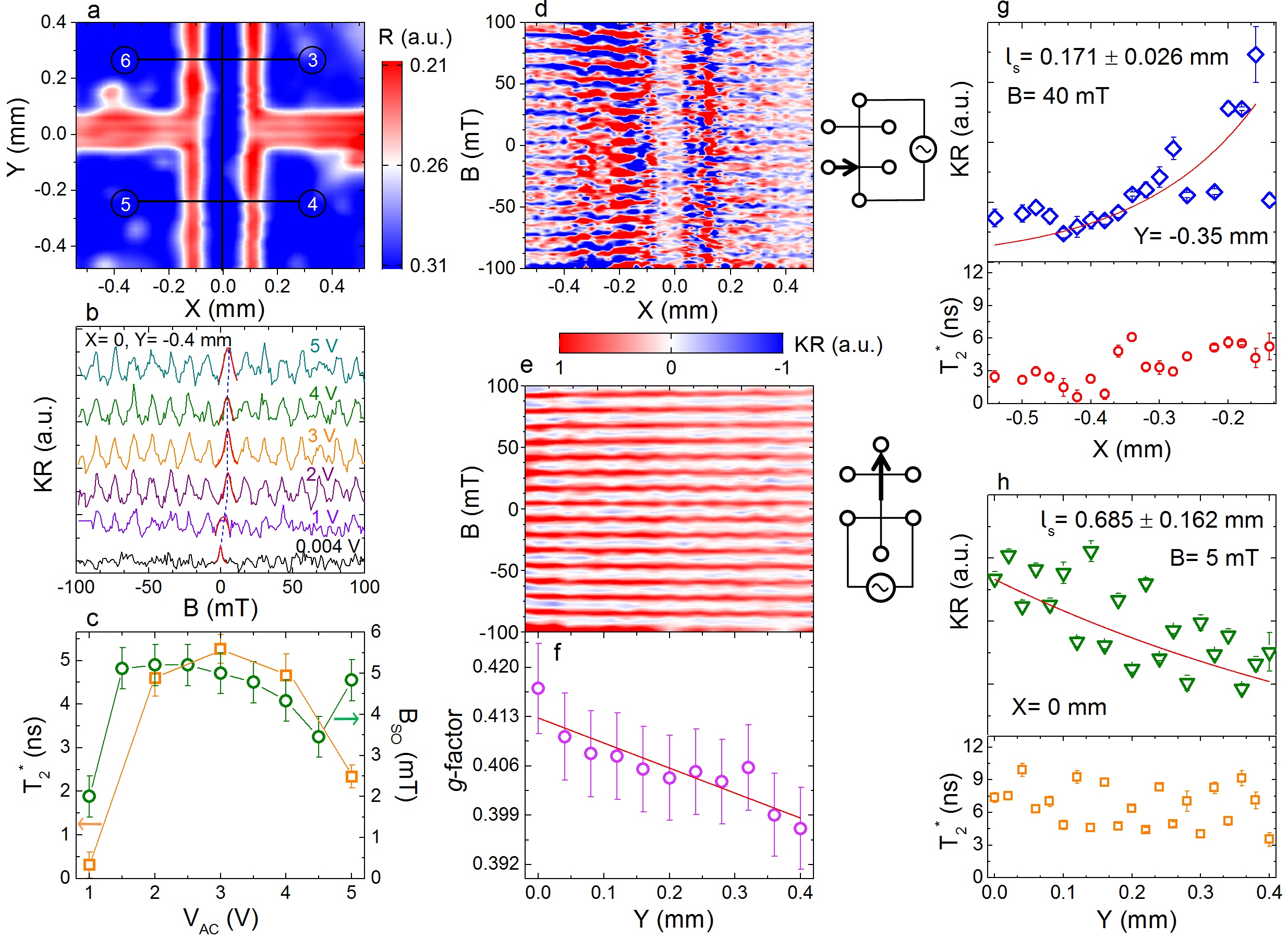}
\caption{Current-induced spin polarization - Transverse configuration. (a) Reflectivity map of the device. (b) B scans of KR at ($x$,$y$) = (0,-0.4) mm applying different $V_{AC}$ in contacts 5-4. The dashed line is a guide to the eyes for the zero-field resonance position at $B=-B_{SO}$. The red line is a Hanle model from where $T_{2}^{*}$ (squares) and $B_{SO}$ (circles) were extracted and plotted in (c). Transverse spin drift along the (d) ($x$,-0.35) and (e) (0,$y$) mm axis with $V_{AC}$= 3 V. The Hall bars display the probe sweeping direction (arrows) and the contacts used for $V_{AC}$ application. (f) Variation of the electron g-factor from the change of the oscillation period in (e). Amplitude of the current-induced spin polarization and spin coherence transported along (g) $x$ and (h) $y$ extracted from (d) and (e), respectively. The fitting of spin polarization decay (solid line) gives $l_S$ as parameter. All the error bars correspond to the fitting standard error. T = 5 K.}
\label{fig5}
\end{figure*}

Next, we analyze the transverse transport of spin coherence in regions where the electric fields are considerably reduced. Figure~\ref{fig5}(a) shows the reflectivity map of the device displaying the central channel and side voltage probes. The low intensity (red) regions indicate the gaps between the conducting areas, the white lines are the edges of those regions and the solid lines are the possible current paths. The horizontal features at $y$ = 0 mm are the gaps separating the contacts 6-5 and 3-4. 

First, we measured the dependence of the CISP amplitude and coherence on $V_{AC}$ using contacts 5-4 and setting the probing spot away at ($x$,$y$) = (0,-0.4) mm. The result is plotted in Fig.~\ref{fig5}(b). We will focus on the spin coherence time for $B=0$ which can be directly evaluated from the width of the zero-th resonance using a Hanle model: $\theta_{K}=A/[(\omega_{L}T_{2}^{*})^{2}+1]$ with half-width $B_{1/2}=\hbar/(g\mu_{B}T_{2}^{*})$. The dashed line guides the field position of the zero-th peak. There, a constant null total magnetic field, arising from the addition of $\bm{B}$ and ${\bm{B_{SO}}}$, requires the shifting of the Hanle peak towards $B=-B_{SO}$.\cite{kalevich} The fitted curve was plotted on top of the experimental data (red line) and the extracted parameters are presented in Fig.~\ref{fig5}(c). In contrast to the longitudinal case, where the spin polarization is measured in a region with a local current flow, the RSA holds over a large voltage range with a $V_{AC}$ enhancement of the spin coherence reaching 5.27 ns at 3 V. $T_{2}^{*}$ decreases for higher voltages but still remains in the nanoseconds range up to 5 V. Complementary, the current-controlled spin-orbit field attains more than 5 mT at 3 V. This field agrees with the theoretical value $\mean{\bm{B}^*_{SO}}$ calculated for $v_{d}=$ 0.5 nm/ps in Fig.~\ref{fig:Bso}. Nevertheless, the expected drift velocity for the longitudinal configuration should be much larger and equal to $v_{d}=I/(en_sw)=$ 18.1 nm/ps for a current $I=$ 5 mA at $V_{AC}=$ 3 V (see Fig.~\ref{figquatro}) with an electric field of $E=V_L/L=$ 0.9 kV/m. Therefore, we find that $v_{d}$ is almost two orders of magnitude smaller in the transverse regions far from the current flow in comparison with longitudinal sections (inside the flow).

The device geometry allows us to examine the variation of the spin coherence transport as a function of the width of the region that connects the current path to the transverse zones. For instance, when $V_{AC}$ is applied along $x$, the transverse region along $y$ consists of the central channel. Reciprocally, when $V_{AC}$ is applied along $y$, the transverse transport will be into the lateral arms of the Hall bar connected to the current path by narrow bridges giving a strong drift constriction in the $x$ axis. We fixed $V_{AC}$= 3 V for the largest spin-orbit field, CISP amplitude, and spin coherence time.

In Fig.~\ref{fig5}(d), we applied $V_{AC}$ in the $y$ axis (contacts 1-2) and scanned the probe position in the $x$ axis (at constant $y=$ -0.35 mm) from the left to the right side arms (5-4) passing by the central channel also. The $x$-scan thus allows us to map simultaneously the longitudinal and transverse transport. Inside the central channel, we reproduced the data from Fig.~\ref{figquatro} where the high current leads to low spin polarization due to heating (white shade from $x=$ 0 to $\pm$0.1 mm). 

Entering the transverse side regions, the polarization was observed in long ranges of $\Delta x=\pm$0.5 mm. In Fig.~\ref{fig5}(e), we used contacts 5-4 approximately 250 $\mu$m away from the measuring spots to emulate a nonlocal geometry. We applied $V_{AC}$ in the $x$ axis and scanned the probe position in the $y$ axis (at $x=$ 0 along the central channel) while sweeping the magnetic field. The CISP transverse transport was observed again in a scan of about $\Delta y=$ 0.5 mm. Nevertheless, the CISP decay now became visibly weaker with distance. 

A remarkable result is the change of the $\Delta B$ period in the RSA pattern observed for both $x$ and $y$ scans. In Fig.~\ref{fig5}(e), this effect is seen by the shift of the outer resonances towards higher fields as a function of $y$. On the other side, the zero-th resonance position seems to remain constant given by $V_{AC}$. Our experimental technique separates the influence of the $B_{SO}$ and $g$-factor changes: While the first behavior depends exclusively on the $g$ factor, the second is related only to $B_{SO}$ [see also Figs.~\ref{fig5}(b) and (c)]. The spatial dependence of $g$ is displayed in Fig.~\ref{fig5}(f). The modification of the g factor by an in-plane electric field was reported recently. For bulk InGaAs epilayers,\cite{kovac} they measured $\Delta g=$ 0.0053 and found that $g$ increases with the drift velocity. For [110] oriented QWs,\cite{chen} the electrical variation of the $g$ tensor was also found to depend on the external magnetic field and to increase with the current.\cite{chen} Our data agrees with those reports as the $g$ factor decreases for electrons that drifted far from the current path. Nevertheless, we reached a variation of $\Delta g=$ 0.02 which could indicate a larger change on the drift velocity because $\Delta g\propto v^2_d$.\cite{kovac}  

We noted that, at the longest transverse distance from the current path ($y=$0.4 mm), we recovered the $g$-factor value measured at $v_d=0$ in Fig.~\ref{fig3}(c). However, the data does not show clearly the expected variation of $B_{SO}$ with $y$ (or $v_d$). This may be explained considering the additional experimental difficulty originated from the fact that the variation with the drift is smaller for $B_{SO}$ than for $g$, since $B_{SO}$ depends only linearly on $v_d$ (see Sec. III C).

Furthermore, the spin transport extension can be inferred from the KR dependence on the spatial parameter ($y$ or $x$) estimated by $exp(-y/l_s)$ where $l_S$ is the CISP transport length. Figures~\ref{fig5}(g) and 5(h) show the KR amplitude and coherence time following a resonance peak in (d) and (e), respectively. The lengths obtained from the exponential decay fitting (solid line) of the spin polarization are $l_S=$ 0.171 mm for scans along $x$ and 0.685 mm for $y$. Figure~\ref{fig5}(h) displays the surprising result that the CISP transverse drift can drive constant spin coherence of about 6 ns by almost half a millimeter. Furthermore, for the $x$-scan, the spin coherence is lost from a value close to 5 ns to around 1 ns in a similar displacement. Those values for $l_S$ and $T_{2}^{*}$ are independent of the field resonance chosen for the analysis within the experimental error (see SM-D).\cite{SM} The measured $l_S$ asymmetry is likely a result of the device geometry. The anisotropy of $B_{SO}$ related to the orientation of the Hall bar with the crystallographic axes was not experimentally evaluated. Our calculation indicated that both in-plane orientations are equivalent (see discussion for Fig.~\ref{fig:Bso}).

\subsection{Nonlocal charge transport}

In this section, we investigate the characteristic length for transverse charge transport in our device. The observed long spin transport length requires not only large spin-orbit coefficients but also needs a large mean free path for charge diffusion. In the clean system, the large charge diffusion will extend the regions where the current induces spin polarization towards transverse directions. On the other side, the large calculated spin-orbit coefficients will then enhance and maximize the efficiency of the polarization generation. We also explored if the large spin drift of the current-induced spin polarization can inversely produce a charge current in close contacts. In the spin Hall effect regime, it is expected that long spin diffusion leads to nonlocal charge transport by means of the inverse SHE.\cite{abanin,schwab} In the Rashba-Edelstein frame, an inverse effect was also proposed for clean electron gases when electron-impurity scattering is very weak in a nonlinear regime.\cite{vignale} The nonlocal transport can be characterized by the transresistance $R_{NL}$ defined by the ratio between the nonlocal response voltage $V_{NL}$(in 5-4) and the applied current $i_{L}$ (in 6-3) for a given $V_{AC}$. 

\begin{figure}[h!]
\includegraphics[width=1\columnwidth]{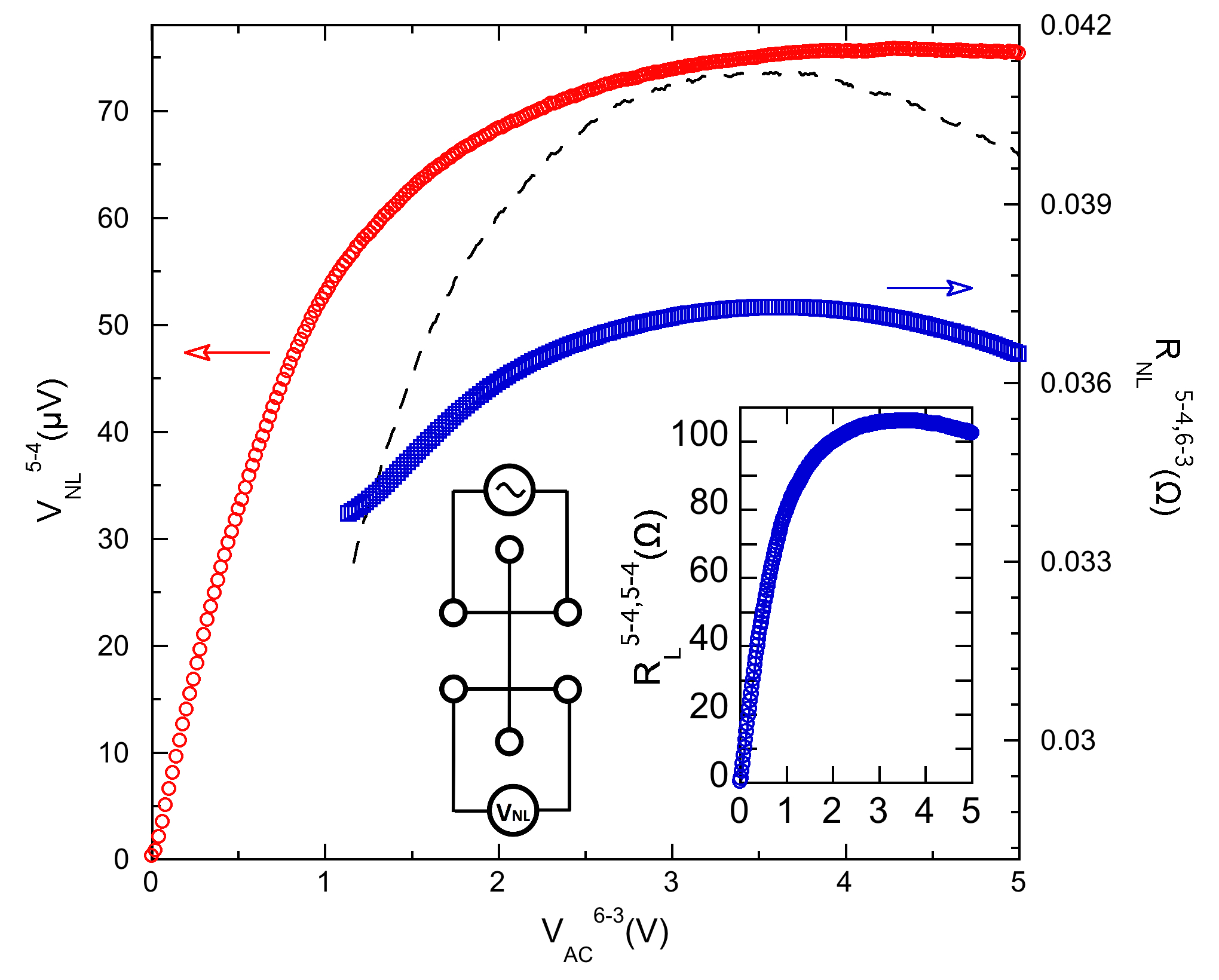}
\caption{\label{fig4} Nonlocal transport. Nonlocal voltage (circles) and transresistance (squares). The inset shows the local resistance and the experimental configuration. The theoretically estimated Ohmic contribution to $R_{NL}$ is shown by the dashed line. $B=$ 0.}
 \label{fig6}
\end{figure}

The experimental configuration for the measurement of the nonlocal voltage is shown as an inset of Fig.~\ref{fig6}. In this configuration, we note that ${\bm{E}}\parallel{\bm{B}}$ and ${\bm{B_{SO}}}\perp\bm{B}$ which could produce an enhancement of spin polarization for fields larger than zero resembling an antisymmetric Hanle curve\cite{kato} or asymmetric if ${\bm{B_{SO}}}$ is not exactly perpendicular to $\bm{B}$.\cite{kuhlen} For $B=$ 0, we observed a nonlocal voltage on the order of tens of $\mu$V increasing with $V_{AC}$ until saturation. Measuring the local current at the source contacts (6-3), we found a maximum $R_{NL}$ of 0.037 $\Omega$ for $V_{AC}=$ 3 V. 

In order to differentiate the contributions from spin mediated transport and classical charge diffusion, the Ohmic component in the nonlocal resistance for a narrow strip can be estimated as $R_{NL}=R_L exp(-y/l_C)$ where $l_C=w/\pi$ and $y=L$ is the distance from the $V_{AC}$ contacts.\cite{pauw,gusevprl} For our device, we obtain a large transport length $l_C=$ 64 $\mu$m implying that the charge mediated mechanism should be important for $R_{NL}$. It also may hinder the spin mediated phenomena even for $y=L=l_S$ due to the large mean free path for charge in the clean system.\cite{abanin} Determining the local resistance, see inset in Fig.~\ref{fig6}, we calculated the expected $R_{NL}$ given by the Ohmic contribution and plotted it with a dashed line. Both curves for $R_{NL}$, experimental and theoretical, present similar peak position but with an amplitude difference of 10\%. Also, the experimental curve is broader keeping the maximum value constant in a larger voltage band. More work is needed to study wire structures in order to isolate the spin and charge-related conductance contributions in all-electrical measurements.

\section{Conclusions and Outlook}

In conclusion, we reported on the realization of current-induced spin polarization in a 2DEG confined in a triple quantum well. We found that the TQW has exceptional properties for the current control of spin-orbit fields, as we calculated and later experimentally observed. The data showed long coherence time for the spin ensemble in the nanoseconds range.

Surprisingly, we observed the drift transport of such current-induced spin polarization over macroscopic distances in a direction transverse to the applied electric field. During the transport, the spin polarization retain its long-lived nanosecond coherence and the polarization amplitude decay was found to be limited by the device geometry. 

The drifting electrons acquire a variation of the Land\'{e} $g$ factor as a function of the velocity controlled by the proximity to the current path. A spin-orbit field was tuned by the applied voltage reaching several mT. The transverse spin transport length was found to be one order of magnitude larger than the Ohmic charge diffusion in the studied configuration. 

The observed long spin transport length can be explained by two material properties in the TQW: large mean free path for charge diffusion in the clean systems, and large spin-orbit coefficients. Future studies in narrow wire channels are still required to distinguish charge mediated from spin mediated transport in all-electrical measurements. Additional measurements with high spatial resolution are still required to verify the calculated proximity of the left and right subbands to the crossed persistent spin helix regime.

This report in a macroscopic device illuminates a path for practical applications using other complex material systems including ferromagnet/semiconductor hybrids and metallic and magnetic thin films.\cite{crooker,borge,wadley} The presented experimental method may be relevant for electrical switching of the direct and inverse spin Hall and spin galvanic effects.\cite{ganichevgal,burkov,skinner}

\section*{Acknowledgments}

F.G.G.H. acknowledges financial support from Grants No. 2009/15007-5, No. 2013/03450-7, and No. 2014/25981-7 and 2015/16191-5 of the S\~{a}o Paulo Research Foundation (FAPESP). G.J.F. acknowledges the financial support from CNPq, CAPES, and FAPEMIG, and thanks Jiyong Fu for helpful discussions. All measurements were done in the LNMS at DFMT-IFUSP.

\newpage 

\section*{Supplemental Material}

\section*{A. Magnetoresistance in perpendicular magnetic fields}

Magnetotransport measurements in triple quantum wells are well described in the recent literature. See References [1-3] for further details of transport studies on similar samples. Figure 7 shows the longitudinal ($R_{xx}$) and transverse ($R_{xy}$ - Hall) magnetoresistance in a perpendicular field. The observed oscillations in $R_{xx}$ consist of the interplay between two types of oscillating phenomena. In 2DEGs, Shubnikov-de Hass (SdH) oscillations occur due to a periodic modulation of electron scattering as the Landau levels consecutively pass through the Fermi level. Additionally, in quantum wells with two or more occupied subbands, the magnetoresistance exhibits another type of oscillations, the so-called magnetointersubband (MIS) oscillations. MIS oscillations occur because of a modulation of the probability of transitions between the Landau levels belonging to different subbands which is periodic in field. The MIS oscillation peaks correspond to the maximal scattering of electrons between the Landau levels when the subband separation equals a multiple of the cyclotron energy. For our high density 2DEG, we can measure only high Landau levels for moderate fields. The quantization of $R_{xy}$ associated with the minima in $R_{xx}$ is also displayed in Figure 7. 

\begin{figure}[h!]
\includegraphics[width=0.9\columnwidth]{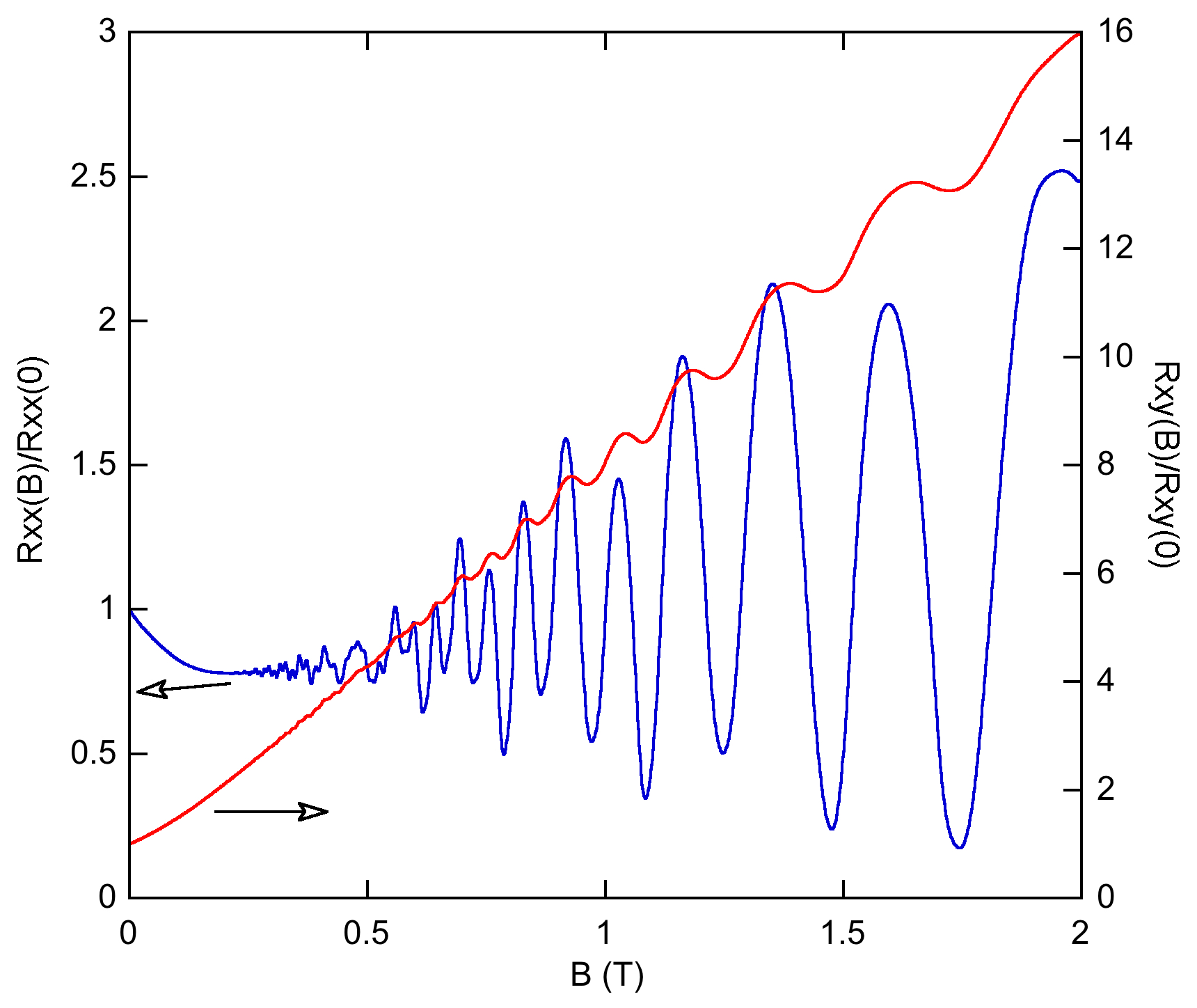}
\caption {Normalized longitudinal ($R_{xx}$) and Hall ($R_{xy}$) resistances with perpendicular field. T = 1.2 K.}
\end{figure}

\begin{figure}[h!]
\includegraphics[width=1\columnwidth]{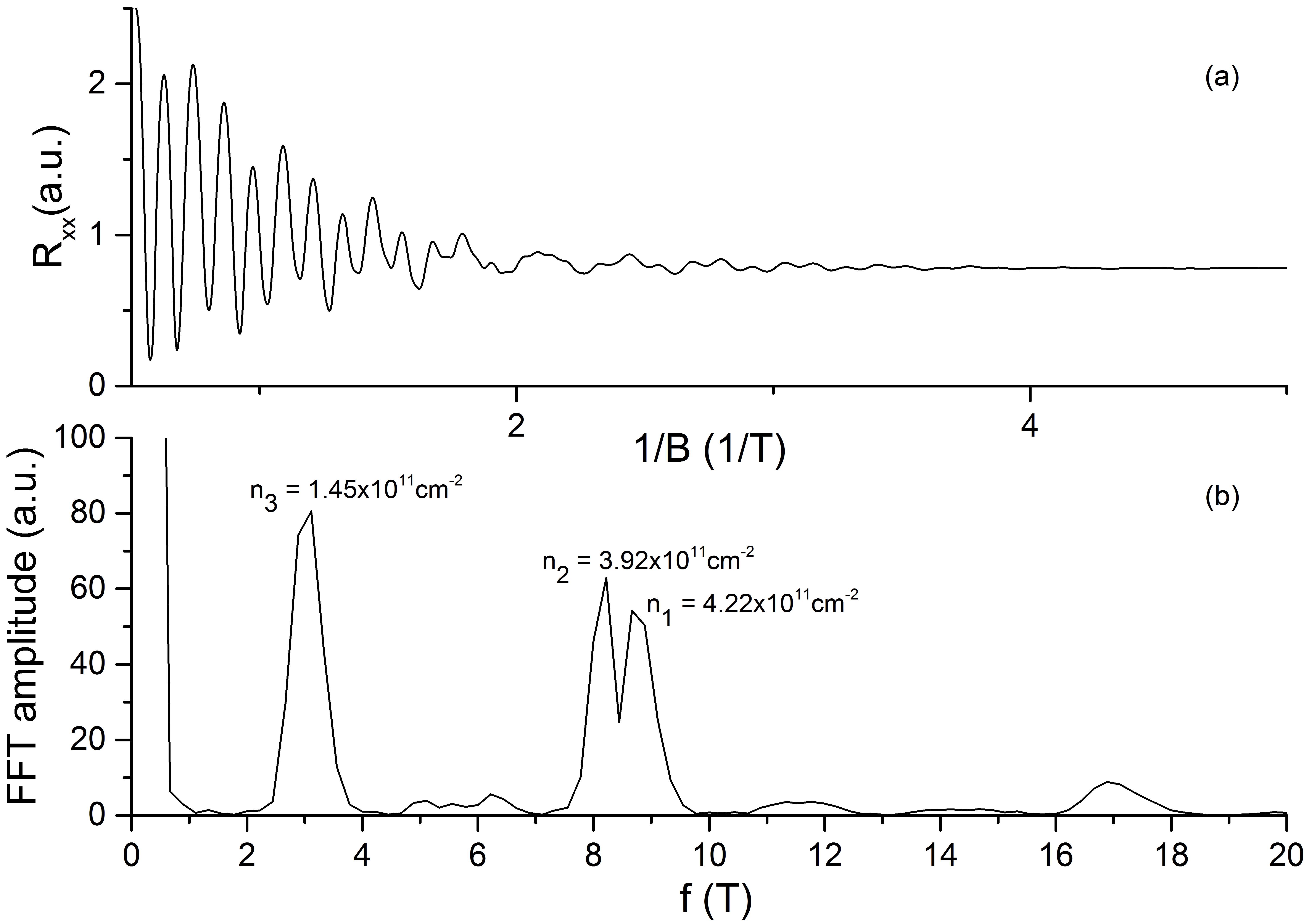}
\caption {(a) SdH oscillations: $R_{xx}$ as function of 1/B, (b) Fourier transform of (a).}
\end{figure}

The SdH oscillations are periodic in 1/B (see Figure 8(a)). From that periodicity, one can obtain the subbands density $n_i$ according to $1/f_i=(2e)/(hn_i)$ where $f_i$ is the frequency of the oscillation in the 1/B plot. In Figure 8(b), the Fourier transform results in 3 large peaks arising from the 3 subbands. The peak with lower frequency corresponds to the lower density in the third subband and the other two correspond to similar densities in the first and second subband. The first two-subbands are almost degenerate ($\Delta_{21}\sim$ 0) as confirmed by our calculation in the main text. We found that $\Delta_{21}$ is very sensitive to the width of the lateral wells (increasing rapidly for smaller widths). 

\section*{B. Wavelength dependence of the time-resolved spin dynamics}

Time-resolved Kerr rotation (TRKR) was used to study the two-dimensional electron gas (2DEG) spin dynamics. Measurements were performed with and without external magnetic fields in order to characterize the ensemble coherence time ($T_{2}^{*}$) and Land\'{e} $g$-factor ($g$) dependence on excitation wavelength ($\lambda$). Figure 9(a) shows TRKR scans for different wavelength at B=0. We found a change from an initial positive spin polarization to negative followed by an exponential decay. The TRKR amplitude at zero time delay ($\Delta t$) is plotted in Figure 9(b). The scans from 817 to 819 nm showed a positive component at shorter delay times but away from $\Delta t=0$. This lineshape was previously associated with the electron spin dynamics in a GaAs/AlGaAs heterojunction system containing a high-mobility 2DEG.\cite{rizoSM}

\begin{figure}[h!]
\includegraphics[width=1\columnwidth]{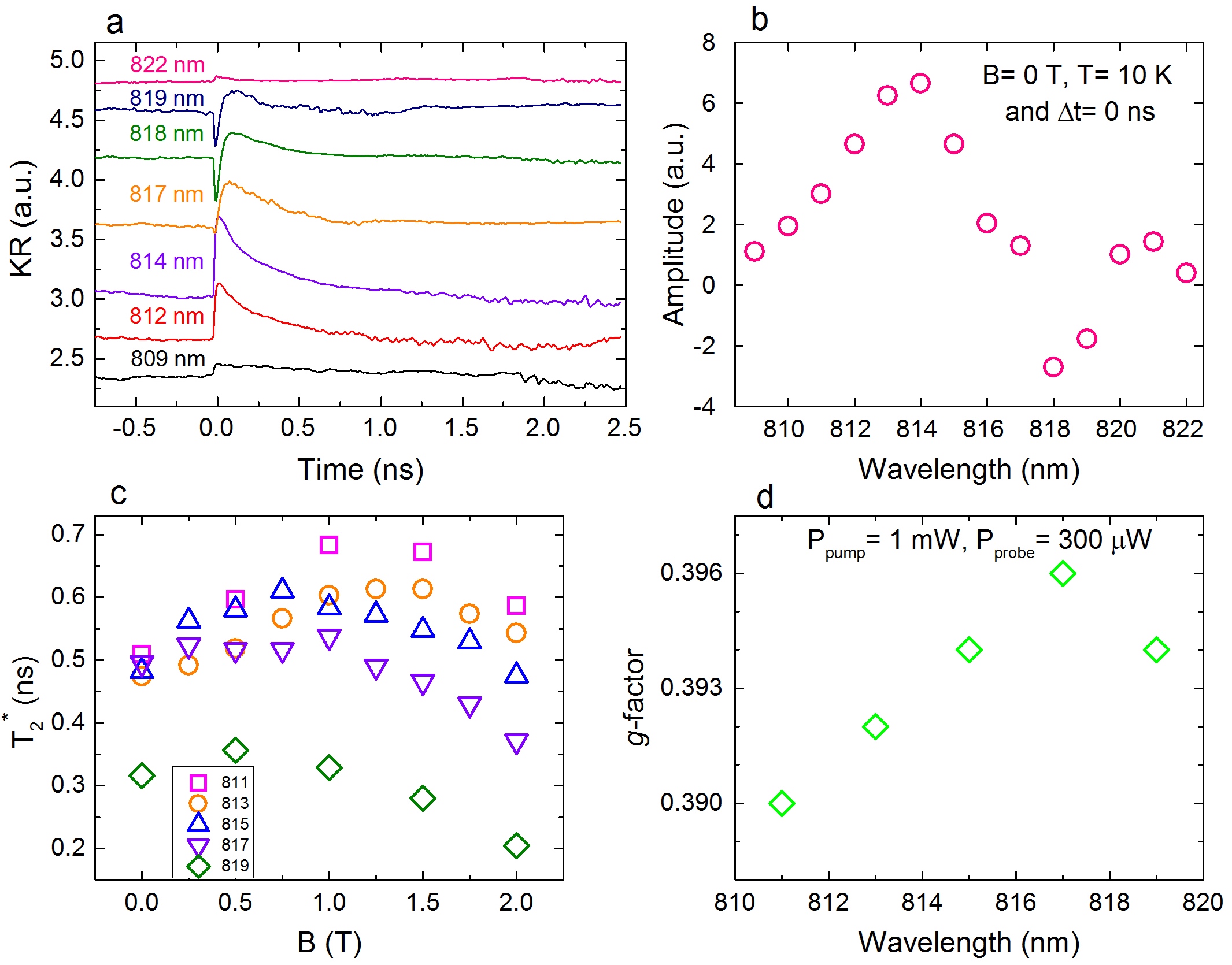}
\caption{For B = 0: (a) Time-resolved KR signal for different wavelengths, (b) Amplitude at time delay $\Delta$t = 0 extracted from (a) plotted as function of wavelength. For B $\neq$ 0: (c) Magnetic field dependence of T$_{2}^{*}$ extracted by fitting the oscillatory part of the KR signal measured for different wavelengths, (d) Electron $g$-factor obtained from the linear dependence of $\omega_{L}$ on the applied magnetic field for different wavelengths. Pump/probe power of 1 mW/300 $\mu$W and T = 10 K.}
\end{figure}

With the application of a external transverse magnetic field, TRKR oscillations are observed arising from the precession of coherently excited electron spins about the in-plane field. To obtain the spin coherence time, the evolution of the Kerr rotation angle can be described by an exponentially damped harmonic:
\begin{equation}
\theta_{K}(\Delta t) = A \exp(-\Delta t/T_{2}^{*})\cos(\omega_{L}\Delta t + \phi)
\end{equation}
where $A$ is the initial spin polarization build-up by the pump, $\phi$ is the oscillation phase, and $\omega_{L}=g\mu_{B}$B/$\hbar$ is the Larmor frequency with magnetic field B, electron $g$-factor (absolute value) $g$, Bohr magneton $\mu_{B}$, and reduced Planck's constant $\hbar$. 

The extracted magnetic field dependence of $\omega_{L}$ and T$_{2}^{*}$ is shown in Fig. 9(c) and (d). We found $T_{2}^{*}$ in the 0.5 ns range decreasing for fields larger than 1 T. Also, the electron g-factor presents a variation with the wavelength (from 811 to 819 nm) of 0.006 having a maximum at 817 nm.

\section*{C. Wavelength and temperature dependence of the optically-controlled CISP}

In the same wavelength range of section A, we investigated the amplitude dependence for the RSA pattern formation at a fixed voltage. The data is shown in Figure 10. 

\begin{figure}[h!]
\includegraphics[width=0.5\columnwidth]{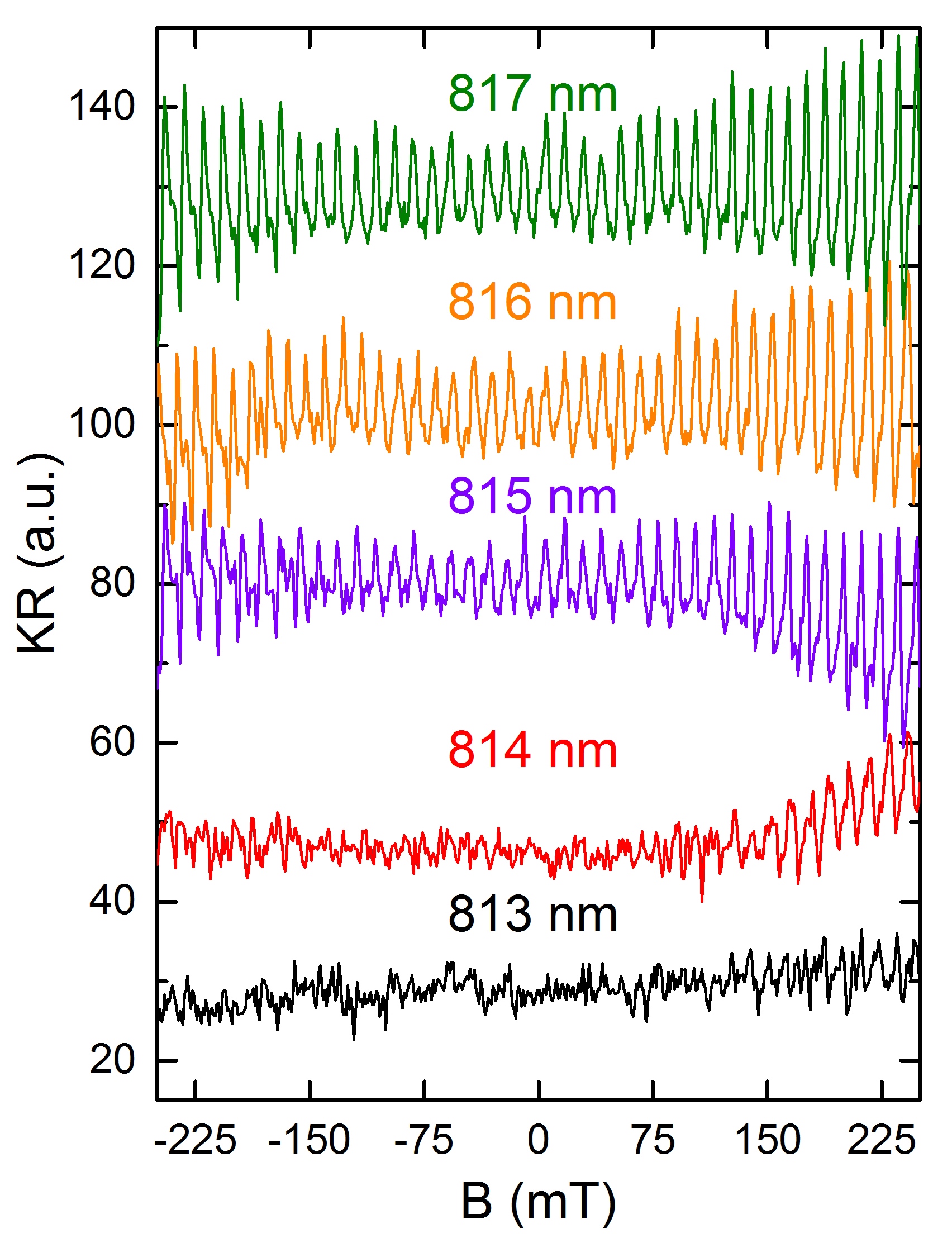}
\caption {Magnetic field scans of the KR signal measured for different probe wavelengths at a fixed power of 300 $\mu$W. $V_{AC}$ = 3 V and T = 1.2 K}
\end{figure}

We found a strong dependence on $\lambda$ with maximum signal at 817 nm. For the spin transport studies, we chose this wavelength as it strongly shows the current effect on the spin coherence added to the results in section A. We fixed $\lambda=$ 817 nm for the following discussions.

\begin{figure}[h!]
\includegraphics[width=1\columnwidth]{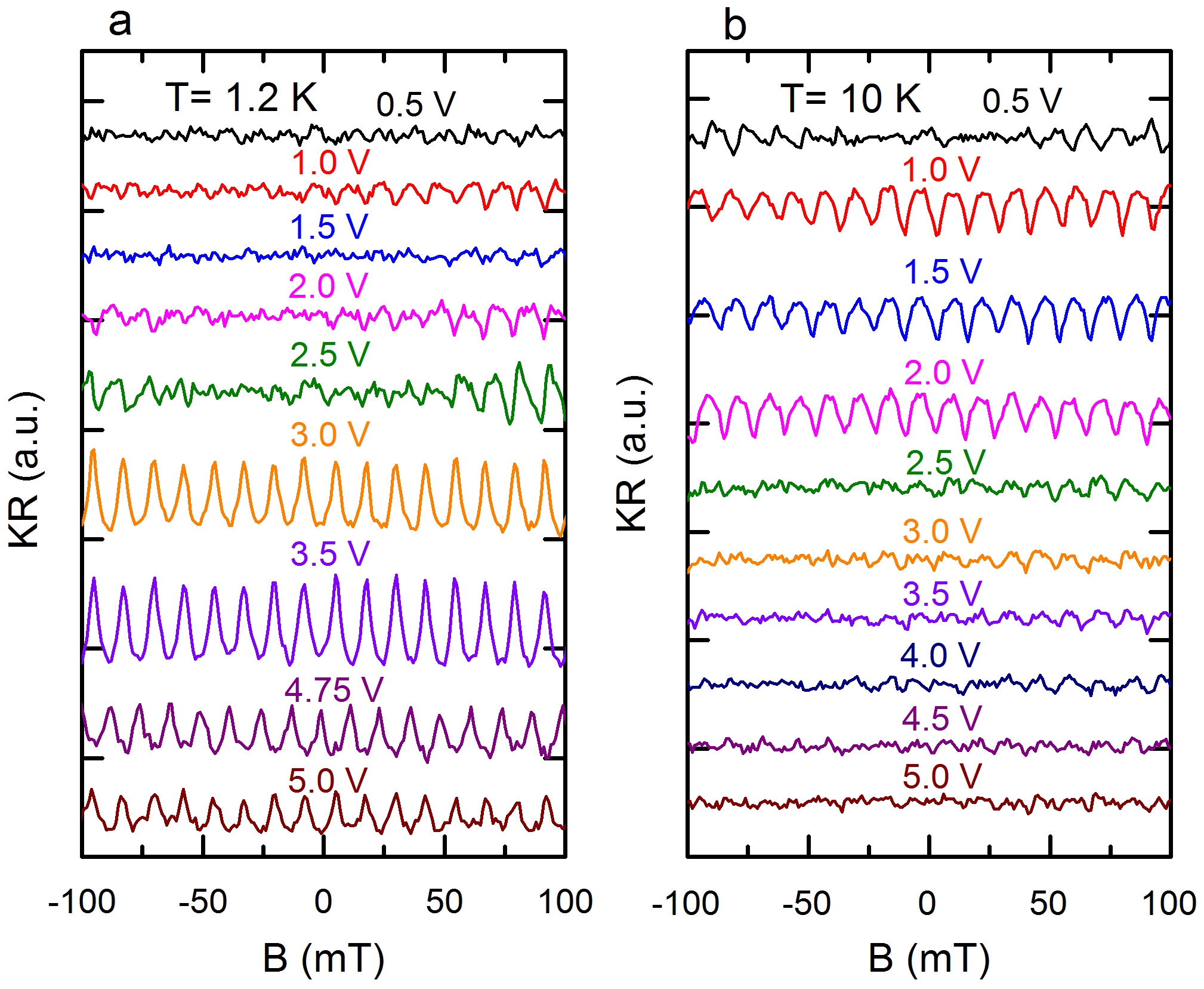}
\caption {Magnetic field scans of the KR signal measured for different applied voltages at T = (a) 1.2 K and (b) 10 K.}
\end{figure}

Next, we explore the temperature influence on the formation of the RSA pattern as function of the applied voltage. We compared the data at 1.2 and 10 K in Figure 11. Clearly, the application of high voltages induces heating which in turn results in strong spin decoherence. This unwanted effect can be delayed by lowering the sample temperature. For T = 1.2 K, the spin polarization amplitude increases with the applied voltage reaching a maximum at 3.5 V followed by a decrease. Furthermore, the resonances have a larger amplitude for larger fields, for example at 2.5 V. This is a known indication of long coherence time for the hole spins involved in the generation of the electron spin coherence.\cite{yugovaSM} For T = 10 K, we cannot observe any current-induced spin coherence beyond 2.0 V.  Also note the drastic change of the resonances width related to loss of spin coherence when increasing the temperature. 

\section*{D. Field dependence of the transport length for spin coherence}

\begin{figure}[h!]
\includegraphics[width=1\columnwidth]{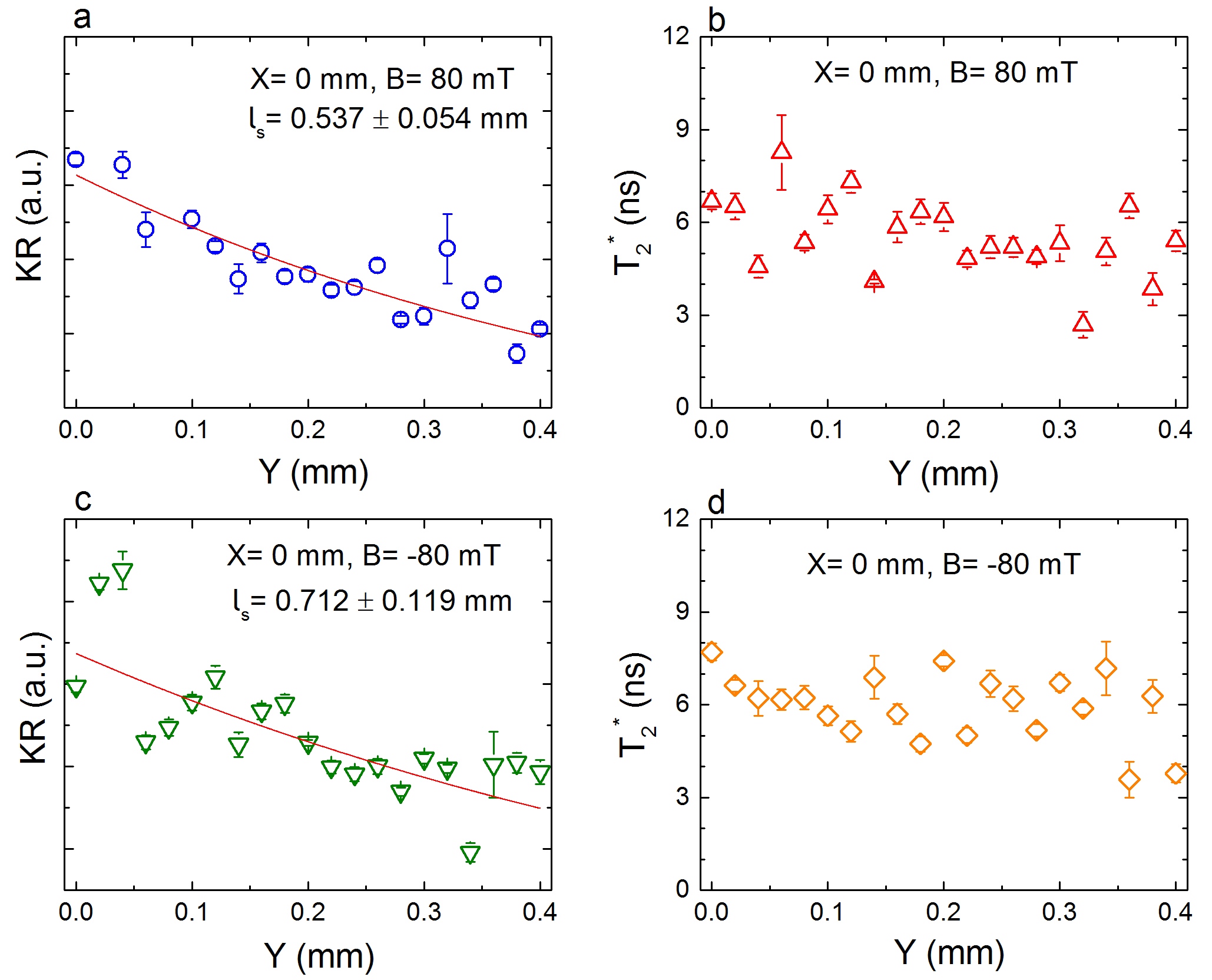}
\caption {KR amplitude and coherence time as function of y for different magnetic fields.}
\end{figure}

\begin{figure}[h!]
\includegraphics[width=1\columnwidth]{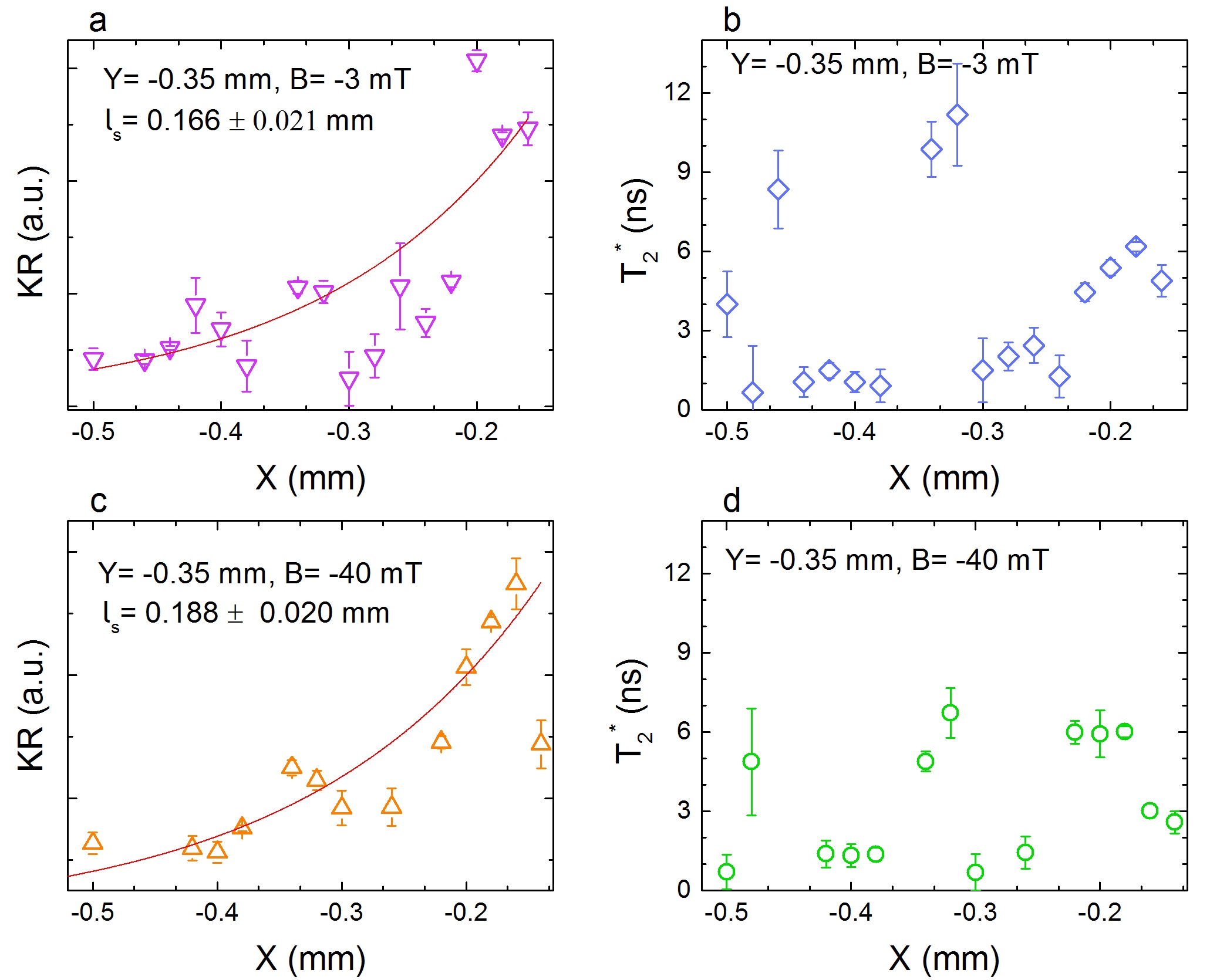}
\caption{KR amplitude and coherence time as function of x for different magnetic fields.}
\end{figure}

In the main article text, we showed the spin transport length $l_S$ extracted from the decay of a single resonance peak at a given magnetic field. This is a condition of approximately constant field since the peak shifts due to the period modification with the $g$-factor variation in space. 

Figures 12 and 13 show fittings, similar to the Figure 5 in the main text, for different conditions of high and low magnetic field as well as different field polarity. All the results are consistent within the experimental error. We conclude that the obtained $l_S$ is robust and the long spin coherence is maintained for mT fields (in agreement with Figure 9(c)).

\end{document}